\documentclass[11pt,a4paper]{article}

\usepackage[T1]{fontenc}
\usepackage{newtxtext,newtxmath}
\usepackage[scaled=0.92]{helvet}
\usepackage{geometry}
\usepackage{microtype}
\usepackage{setspace}
\usepackage[super,sort&compress]{natbib}
\usepackage[colorlinks=true,allcolors=black,urlcolor=NatureBlue]{hyperref}
\usepackage{graphicx}
\usepackage{float}
\usepackage{booktabs}
\usepackage{tabularx}
\usepackage{array}
\usepackage{ragged2e}
\usepackage{colortbl}
\usepackage{pdflscape}
\usepackage{afterpage}
\usepackage{amsmath}
\usepackage{tikz}
\usetikzlibrary{arrows.meta,positioning,calc,fit}
\usepackage[most]{tcolorbox}
\usepackage{enumitem}
\usepackage{caption}
\usepackage{titlesec}
\usepackage{fancyhdr}
\usepackage{lineno}
\usepackage{xcolor}
\usepackage{url}

\definecolor{NatureBlue}{HTML}{0B5A75}
\definecolor{NatureTeal}{HTML}{167D78}
\definecolor{NatureGold}{HTML}{B47B18}
\definecolor{NatureRed}{HTML}{A13A32}
\definecolor{NatureInk}{HTML}{20252A}
\definecolor{NatureMid}{HTML}{66717A}
\definecolor{NaturePale}{HTML}{EFF4F5}
\definecolor{NatureWarm}{HTML}{F8F1E5}
\definecolor{NatureRose}{HTML}{F8EBE8}
\definecolor{NatureLine}{HTML}{CBD5D9}

\usepackage{booktabs,tabularx,array,ragged2e,xcolor,colortbl,caption}
\definecolor{RelTableHead}{HTML}{EDF1F3}
\definecolor{RelTableLLM}{HTML}{F8F4EA}
\newcolumntype{Z}[1]{>{\hsize=#1\hsize\linewidth=\hsize\RaggedRight\hyphenpenalty=10000\exhyphenpenalty=10000\arraybackslash}X}
\newcommand{\RelTableStyle}{%
  \fontfamily{qhv}\fontsize{9}{11.5}\selectfont
  \setlength{\tabcolsep}{2pt}\setlength{\extrarowheight}{2pt}%
  \renewcommand{\arraystretch}{1.12}}
\newcommand{\RelTableHeader}{%
  \rowcolor{RelTableHead}
  \textbf{Dimension} & \textbf{Coin toss} & \textbf{IEC 61508 safety software} &
  \textbf{DL classifier} & \textbf{Autonomous vehicle} &
  \cellcolor{RelTableLLM}\textbf{LLM service}\\[3pt]}

\hypersetup{
  pdftitle={Reliability is a user-specific property of large language models},
  pdfauthor={Xingyu Zhao},
}

\titleformat{\section}{\sffamily\bfseries\Large\color{NatureInk}}{}{0pt}{}
\titleformat{\subsection}{\sffamily\bfseries\normalsize\color{NatureBlue}}{}{0pt}{}
\titlespacing*{\section}{0pt}{22pt}{8pt}
\titlespacing*{\subsection}{0pt}{14pt}{3pt}

\renewcommand{\headrulewidth}{0.35pt}
\renewcommand{\headrule}{\hbox to\headwidth{\color{NatureLine}\leaders\hrule height \headrulewidth\hfill}}

\modulolinenumbers[5]
\newcolumntype{Y}{>{\RaggedRight\arraybackslash}X}
\newcolumntype{P}[1]{>{\RaggedRight\arraybackslash}p{#1}}
\newcommand{\keyclaim}[2]{%
  \begin{tcolorbox}[
    enhanced,colback=NaturePale,colframe=NatureBlue,
    boxrule=0pt,borderline west={2.4pt}{0pt}{NatureBlue},
    arc=0pt,left=9pt,right=9pt,top=7pt,bottom=7pt,
    before skip=11pt,after skip=11pt]
    {\sffamily\bfseries\color{NatureBlue}#1}\quad #2
  \end{tcolorbox}}

\begin{document}

\begin{flushleft}
{\sffamily\bfseries\small\color{NatureBlue} PERSPECTIVE}\par
\vspace{7pt}
{\sffamily\bfseries\fontsize{25}{29}\selectfont\color{NatureInk}
Reliability is a user-specific property of large language models\par}
\vspace{12pt}
{\large Xingyu Zhao$^{1,2,*}$}\par
\vspace{5pt}
{\small\itshape $^{1}$Wuhan University, Wuhan, China.\par
\small\itshape $^{2}$University of Warwick, Coventry, United Kingdom.\par
$^{*}$e-mail: xingyu.zhao@whu.edu.cn}
\end{flushleft}

\vspace{8pt}
\noindent{\sffamily\bfseries\color{NatureInk} Abstract}\par
\noindent\textbf{Public benchmarks support comparisons of large language model (LLM) capabilities, yet users can experience the same highly ranked model differently. Drawing on traditional software reliability, we argue that the reliability of an LLM service should be assessed for specified users and conditions of use. This requires more than a capability score: evaluation evidence must support a claim about failure during future operation. We distinguish testing to discover faults from testing to assess reliability and identify ten modelling considerations behind such claims. We propose a six-workstream research roadmap linking claim specification, dynamic operational profiles, graded outcomes and noisy judgements, local probabilistic robustness, interaction dependence, and statistical inference with reassessment. Local probabilistic robustness addresses the behavioural variation hidden when different concrete inputs are treated as the same task. Published findings and an illustrative customer-support example motivate the roadmap; its integrated validation remains an open research task. Reliability claims complement, but do not replace, separate safety and security arguments. Capability scores should therefore be complemented by versioned operational claims that state whose requirements are covered, under which conditions, and with what uncertainty.}

\vspace{12pt}
\noindent\rule{\textwidth}{0.55pt}

Public benchmarks are valuable for capability assessment, model comparison, regression testing and development. MMLU, BIG-bench, HELM, MT-Bench and Chatbot Arena help researchers and companies evaluate models and track progress\citep{hendrycks2021mmlu,srivastava2023bigbench,liang2023helm,zheng2023judge,chiang2024arena}. Yet users can experience the same highly ranked model differently: one receives consistently useful assistance, while another encounters frequent errors, inappropriate refusals or inconsistent advice. Existing research has identified several possible explanations, including benchmark contamination, limited task coverage, prompt sensitivity and aggregation choices\citep{singh2025leaderboard,sclar2024sensitivity}. These limitations matter, but addressing them alone does not resolve a more fundamental issue: a benchmark score and an operational reliability claim answer different questions. A benchmark score summarizes behaviour on selected tasks under a scoring procedure; reliability concerns the probability of providing the required service without failure during future use under stated conditions. The deeper challenge is therefore to establish what claim the evaluation evidence supports, for which users and conditions of use.

A test suite designed to reveal faults by triggering failures does not automatically estimate how often those failures occur in operation. Representative acceptance testing may reveal fewer previously unknown faults while providing stronger evidence about reliability under intended use. A benchmark need not have been designed to supply all these objects. When its scope and sampling match the intended use, or a justified argument connects them, benchmark observations may contribute to reliability assessment. The research problem is to make that connection explicit and testable.

Reliability should be distinguished from safety and security. Reliability concerns sustained performance of required service under stated conditions; safety concerns freedom from intolerable harm; security concerns protection of systems and assets against threats, including unauthorized and adversarial actions \citep{avizienis2004taxonomy,bloomfieldRushbyWhereAIWrong2024}. They interact: a security breach can cause service failure and harm, while frequent harmless errors can reduce reliability without creating a safety hazard. This article develops operational reliability claims, rather than on safety, security, or other properties\footnote{These can support safety and security arguments, but require additional hazard and threat analyses to do so.}.

Traditional software reliability supplies the vocabulary and statistical discipline for this purpose: testing objectives, operational profile (OP), demand and exposure, stochastic failure process, oracle, dependence, estimator, uncertainty and validity over time \citep{littlewood2000roadmap,musa1993op,lyu1996handbook}. Its lessons remain relevant: fault counts differ from failure probabilities; fault finding and operational testing support different decisions; independence assumptions can dominate the evidence \citep{adams1984optimizing,salako2023unnecessity}. Our position is that \emph{LLM reliability should be, grounded in traditional software reliability engineering, assessed for specified users and conditions of use}. This requires more than a user-specific OP \citep{chakherlou2026hip}. Three connected questions organize the ten modelling considerations (cf. later Table~\ref{tab:comparison}) and guide the proposed roadmap.

\textbf{What are we assessing, and how is it measured?} Define the required service, the failure measure and the unit of assessment, such as a task, a conversation or service hours. Outcomes may be graded, failures differ in severity, and the people or models judging them may be noisy or biased.

\textbf{How should operational use and system behaviour be modelled?} Describe task frequencies and variation within tasks, including task-preserving changes in wording or formatting. Stochastic generation, conversation history and service state also affect failures and can make observations dependent.

\textbf{What level of confidence does the statistical evidence justify?} Connect the sampling design and analysis to intended use, quantify uncertainty, and state when changes in the service, its use or its measurement require reassessment.

These questions complement contextual and behavioural evaluation and research on personalized alignment; they do not suggest that existing evaluation ignores users or conditions of use.\citep{liang2023helm,ribeiro2020checklist,kirk2024bounds} The additional emphasis is on the inference from observed outcomes to a bounded claim about future service. This Perspective proposes research questions and candidate methods, not a validated end-to-end method. Published empirical findings motivate the agenda; the customer-support example illustrates it without claiming new deployment data.

\section{Two purposes of testing, and the ``5000-year bugs''}

Testing can be classified from different perspectives, e.g., by lifecycle stage, access to internals, and level of automation. From the perspective of purpose, however, two missions organise the evidence: \emph{debug testing} and \emph{acceptance testing}\footnote{Related names include fault-finding, defect, directed, stress or challenge testing for debug-oriented work; and operational, statistical, reliability-demonstration or reliability-acceptance testing for acceptance-oriented work. Terminology varies; the distinction here is the primary decision that the evidence is intended to support.}. Debug testing seeks informative failures so that faults can be diagnosed and removed. It deliberately favours hard, diverse or suspicious cases. Acceptance testing treats the system as if deployed and asks whether evidence supports a stated residual-risk claim under anticipated use.\citep{frankl1998delivered,cotroneo2016relai,guerriero2021operation,zhao2025statistical}

The distinction resembles driving practice and a driving examination. During practice, an instructor deliberately repeats difficult manoeuvres to identify and correct weaknesses. During an examination, an examiner follows a stated protocol to support an acceptance decision about the candidate’s current driving performance under conditions intended to resemble real-world driving\footnote{A driving examination aims to mimic selected aspects of everyday driving, but its routes, duration and scenarios are not necessarily statistically representative of everyday use. The analogy concerns the purpose of testing, not a claim that passing the examination establishes operational reliability.}. The activity is similar, but the purpose and interpretation of the evidence differ. The analogy concerns testing purpose: reliability acceptance additionally requires a justified connection to operational use and statistical inference.

Before generating a test, the assessor should therefore ask which purpose and decision it must support. A suite designed to discover faults (by triggered failures) does not automatically estimate how frequently they will occur in operation. Representative acceptance testing may find fewer novel faults while providing stronger evidence for the delivered reliability (that as if the system-under-testing was deployed in operation). That said, a testing campaign can serve both debug testing and acceptance testing at the same time through carefully designed sampling and appropriate statistical inference\footnote{To use failure-seeking tests for reliability assessment, the sampling strategy must be paired with an estimator that corrects for differences between test-case frequencies and the OP. For example, importance sampling can test failure-prone cases more frequently than they occur in operation, then reweight their outcomes to obtain an unbiased estimate of operational failure probability under the stated sampling assumptions.}, and combining these purposes has been studied in software engineering \citep{cotroneo2016relai,guerriero2021operation}. 

The distinction matters because \textit{a system with many faults can be reliable, whereas a system with few faults can be unreliable}. Adams's study of nine large IBM software products provides the classic illustration. About one third of observed faults belonged to a class with an average activation interval of roughly 5,000 years \citep{adams1984optimizing}. In a simple additive-rate illustration, removing 100 ``5,000-year faults'' reduces the aggregate failure rate by 0.02 failures per year. Removing one fault activated every five years reduces it by 0.2 failures per year, ten times as much. These are only illustrative calculations, not rigorous measurements. Reliability depends on how often faults are activated under use; an ``unweighted bug count'' does not directly support reliability claims.

\keyclaim{Remark 1 | Bug count is not reliability.}{As the example illustrates, a system with many ``5,000-year faults'' may be more reliable than one with only a few ``5-year faults''. What matters is how often faults cause failures under intended use, not their number. Testing should therefore begin with a clear purpose, with test selection and interpretation matched to the claim being made.}

The same distinction clarifies the role of benchmarks. Capability comparisons, regression detection and behavioural discovery are legitimate objectives; suites such as CheckList make behavioural testing particularly explicit\citep{ribeiro2020checklist}. A benchmark may also support acceptance when its evidence is connected to an operational population and failure criterion. Its label alone determines neither purpose. The relevant question is what inference its design supports: a comparative score answers one useful question, while supporting a reliability claim additionally requires explicit failure oracle, a justified link between test sampling and the OP, and statistical inference that quantifies uncertainty.

\keyclaim{Remark 2 | Benchmark scores alone do not establish operational reliability.}{Challenge-oriented benchmarks are closer to debug testing than to reliability acceptance testing: their curated, often difficult tasks expose LLM weaknesses rather than reproduce a specified user's OP. Some benchmarks also seek representative task coverage, but coverage alone does not establish a match to actual use.}

\section{From traditional software to LLMs: reliability is user-specific}

Dependability standards define reliability as the ability of an item to perform as required, without failure, for a given time interval under given conditions \citep{iec60050}. The probabilistic content is familiar for hardware, where random physical variation and degradation motivate distributions of lifetime or time to failure. Software does not wear out in the same sense. A design fault in a frozen deterministic program is activated systematically for a fixed input and state. However software failure occurrence is uncertain to an assessor due to: 1) incomplete knowledge (epistemic uncertainty) of which inputs and conditions trigger failures, and 2) (aleatory) uncertainty about which inputs, states, configurations, and environments will be encountered during future operation, thus necessitating probabilistic treatment \citep{littlewood2000roadmap,strigini_guidelines_1997}.

\begin{tcolorbox}[
  enhanced,colback=white,colframe=NatureInk,
  boxrule=0.7pt,arc=0pt,left=11pt,right=11pt,top=9pt,bottom=9pt,
  before skip=13pt,after skip=13pt,
  title={\sffamily\bfseries Box 1 | A minimal reliability vocabulary},
  coltitle=white,colbacktitle=NatureInk,toptitle=5pt,bottomtitle=5pt]
\small
\textbf{Reliability.} In dependability terminology, the ability to perform as required without failure for a stated time interval under stated conditions \citep{iec60050}. A quantitative claim must specify those requirements, conditions and exposure.

\textbf{Operational profile, $P_{\mathrm{op}}$.} The probability distribution of demands and relevant operating conditions in use. In a discrete demand-based example, $P_{\mathrm{op}}(x)$ is the probability of demand $x$.

\textbf{Failure.} Failure occurs when the system does not provide the required service under the specified conditions. A claim must state which failures it concerns; different consequences can justify separate severity categories.

\textbf{Oracle.} The procedure used to decide whether a failure occurred, using a specification, automated checks or review. An actual failure and the judgement about it are distinct: the procedure can miss failures or incorrectly label acceptable outcomes.

\textbf{Probability of dangerous failure on demand (PFD).}
Following IEC 61508, PFD expresses safety unavailability: the probability that a safety-related system is unable to provide its specified safety function when required\citep{iec61508part4}.

For a simplified software-testing representation, fix the configuration and reset conditions, and let $f(x)$ be the deterministic response to demand $x$. Assume an accurate binary oracle $\mathcal O_F(x,f(x))$, equal to 1 for a dangerous failure on demand and 0 otherwise. Then \cite{strigini_guidelines_1997,littlewood1997stopping,bishop2011formalism}
\begin{equation}
\mathrm{PFD}
=\mathbb E_{X\sim P_{\mathrm{op}}}\!\left[\mathcal O_F(X,f(X))\right]
=\sum_{x\in\mathcal X}\mathcal O_F(x,f(x)) P_{\mathrm{op}}(x)\,.
\label{eq:pfd}
\end{equation}
The OP supplies demand frequencies; the oracle identifies failures; PFD combines them into a probability. This illustrates operational testing, rather than prescribing an IEC calculation. State- or time-dependent availability requires a richer model. The complement $1-\mathrm{PFD}$ concerns successful response to a random next demand (sampled from the OP), not reliability over a time interval.
\end{tcolorbox}

Software reliability is therefore \emph{user-centric}: it depends on how the software is used, not only on the software itself. Even when two users run exactly the same program, one may frequently encounter inputs that trigger a fault, while the other rarely encounters them. Their probabilities of experiencing failure can therefore differ. The OP describes this use: which demands and operating conditions occur, and how often \citep{musa1993op}. Users may also have different requirements, so a reliability assessment must specify both the intended use and what counts as a failure.

\begin{figure}
    \centering
    \includegraphics[width=1\linewidth]{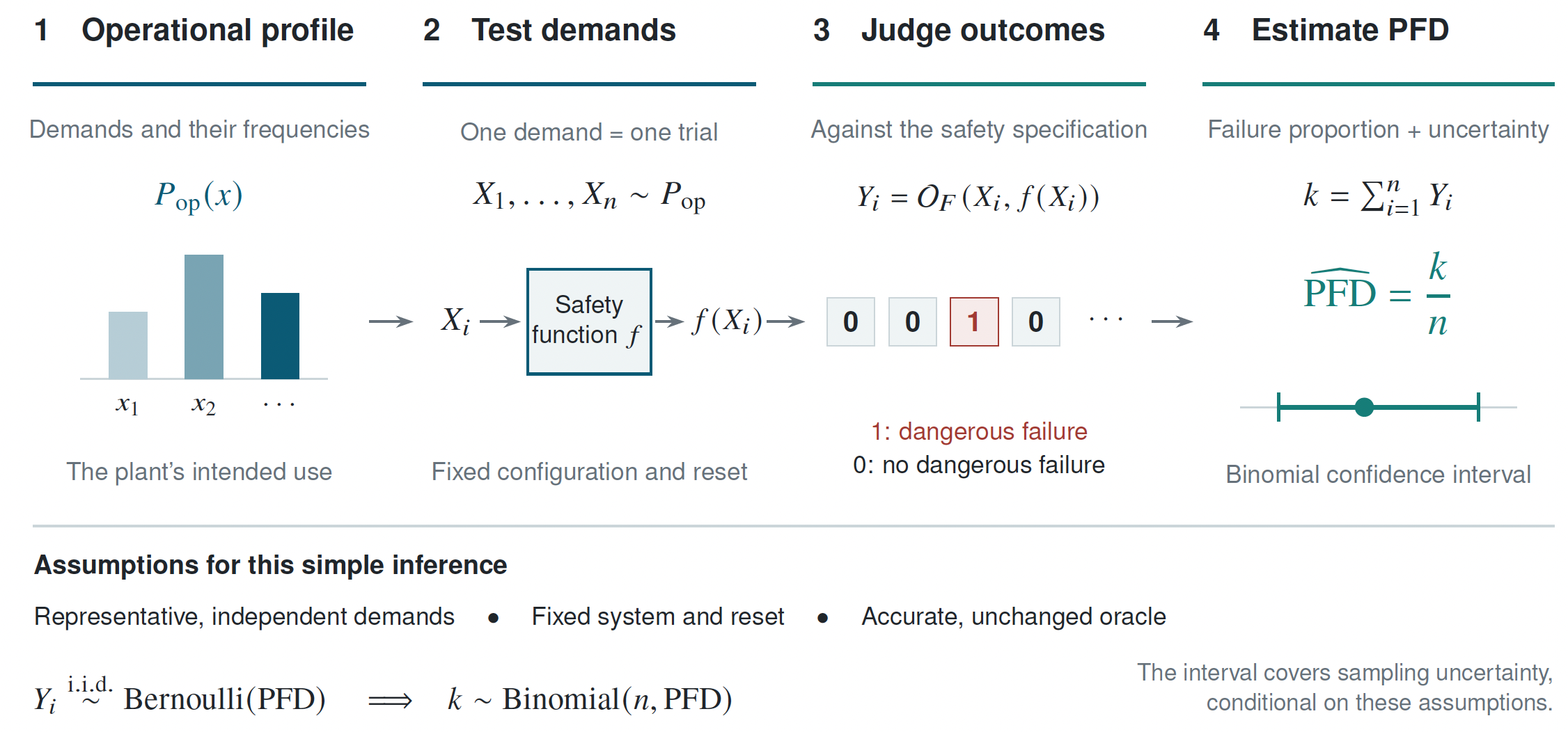}
    \caption{A simple demand-based system reliability assessment of PFD. Demands are sampled from the OP and executed under fixed configuration and reset between demands. An accurate, specification-based oracle identifies dangerous failures. Under the stated assumptions, outcomes are independent Bernoulli trials: the failure proportion estimates the probability of dangerous failure on demand (PFD), and a binomial confidence interval quantifies sampling uncertainty. The illustrated distribution, outcomes and interval are schematic.}
    \label{fig_PFD_inf_chain}
\end{figure}

The familiar demand-based case shows the inferential chain, cf. Fig.\ref{fig_PFD_inf_chain}. Suppose one statistical experiment is one demand on an IEC 61508 safety function, with dangerous failure judged against its specification. For the simplified setting in Box 1, tests draw demands from the plant OP. If successive demands are independent draws from the same OP, the configuration and reset conditions remain fixed, and the oracle is accurate, their binary outcomes can be modelled as independent, identically distributed Bernoulli trials. The sample failure proportion then estimates PFD; a confidence or credible interval states what finite evidence supports. These are assumptions for this simple assessment, not consequences of the PFD definition. For continuous operation, the analogue may be a failure rate per hour, mile or transaction, with Poisson, survival or recurrent-event models rather than Bernoulli trials\citep{iec61508,littlewood1997stopping,miller1992zero,strigini_guidelines_1997,zhao2020operational,min2022recurrent}.

Each link is conditional. A test profile different from the OP can bias the estimate. Oracle error can create false precision. Dependence can make 2,000 observations contain far less information than 2,000 independent trials. A change in software, hardware, configuration or environment can invalidate the claim. Traditional software reliability has methods for most of these complications, including conservative inference under incomplete prior knowledge, uncertain profiles, correlated executions and evidence carried between versions\citep{bishop2011formalism,pietrantuono2020op,salako2023unnecessity,salako2024correlated,littlewood2020replacement,zhao2018perfection}.

\textit{An LLM is still software, so the reliability concepts and inferential chain above remain relevant}. A formal definition grounded in this tradition is first given by HIP-LLM\citep{chakherlou2026hip}: LLM reliability is the probability of failure-free operation over a specified number of future tasks sampled independently from a given OP. The central question therefore remains the same: how likely is the service to meet stated requirements during its intended future use?

The modelling, however, becomes much harder. Tasks may span multiple turns, and their frequencies and forms vary across users and over time. Outcomes may be graded, some requirements involve subjective judgements, and human or model judges may make mistakes. Stochastic generation, conversation history and shared service state complicate independence assumptions. Even task-preserving changes in wording can alter model behaviour, while changes to the service configuration can make earlier evidence less relevant. These difficulties call for more careful modelling, not a different meaning of reliability. The next section examines them through a progression from coin tosses to safety-critical software, deep-learning classifiers, autonomous vehicles and LLMs, using ten modelling considerations to explain why LLM reliability assessment is harder.

\keyclaim{Remark 3 | From traditional software to LLMs: the reliability concepts should remain.}{The modelling, however, becomes much harder. This includes defining intended use and failure, judging outcomes, modelling input variation and dependence across interactions, and justifying statistical claims about future operation from finite test evidence.}

\section{From coin tosses to LLMs: why reliability modelling becomes harder}

Table~\ref{tab:comparison} compares five estimation problems of coin tosses, safety-critical software, deep-learning classifiers, autonomous vehicles and LLM services (columns), along ten dimensions (rows) that must be settled before modelling a reliability: claim, statistical unit, failure and severity, oracle, stochastic process, dependence, OP, local robustness, uncertainty and version validity. The first four define what is assessed and measured; the next four describe operational use and behaviour; the final two govern the confidence and lifetime of the claim. The following discussion follows the table's row order, grouping closely related considerations. These problems are not unique to LLMs, nor does every LLM application require a more complex model than every earlier system. The comparison shows why several difficult modelling decisions arise together in interactive, user-facing LLM services.

\begin{table}[p]
\centering
\caption{\textbf{Reliability-assessment modelling from a coin toss to an LLM service.} \textbf{a}, Claim, measurement and failure process. Each column is illustrative rather than prescriptive; entries must be justified for the actual claim.}
\label{tab:comparison}
\RelTableStyle
\begin{tabularx}{\linewidth}{@{}Z{0.85}Z{0.65}Z{1.05}Z{1.00}Z{1.10}>{\columncolor{RelTableLLM}}Z{1.35}@{}}
\toprule
\RelTableHeader
\midrule
\textbf{Claim / metric} & Probability of heads; probability of all heads in the next $N$ tosses & PFD; probability of failure-free operation over the next $N$ demands & Misclassification probability per input & Crash, fatality or disengagement probability per scenario or trip; event rate per mile or hour & Failure probability or expected severity per user task or conversation; specified-event rate per response, token or hour \\[4pt]
\midrule
\textbf{Statistical unit / exposure} & One toss & One demand on the safety function & One prediction on a sampled input & One scenario or trip; or continuous distance or time & One bounded task, possibly multiround; or cumulative exposure such as service time or number of tokens \\[4pt]
\midrule
\textbf{Failure / severity} & Tails when heads are required; binary & Failure to perform the specified safety function; consequences assessed through hazard analysis & Invalid class or decision; binary, class-weighted or loss-valued & Crash, unsafe manoeuvre, near miss or disengagement; consequence-based severity & Hallucination, unsafe response, privacy loss or unresolved task; ordinal or continuous severity, potentially user-dependent \\[4pt]
\midrule
\textbf{Oracle / observation} & Direct physical observation & Specification, plant state, logs and test oracle & Reference or expert label and task rule; labels may be wrong & Telemetry, event reconstruction, safety rules and review & Rules or human/model judgement of recorded evidence; disagreement and systematic bias are possible \\[4pt]
\midrule
\textbf{Stochastic process} & Bernoulli sequence & Fixed-version Bernoulli demands; a state-dependent sequence if reset is incomplete & Binary failure or categorical outcomes & Bernoulli by scenario or trip; Poisson, NHPP, survival or recurrent-event models by exposure & Binary, ordinal or loss-valued task outcomes; hierarchical, hidden-state or recurrent-event models for interaction \\[4pt]
\bottomrule
\end{tabularx}
\par\smallskip
{\fontsize{8}{10}\selectfont\RaggedRight
A failure probability, an event rate and an expected severity are distinct assessment targets. In the IEC column, PFD concerns dangerous failure of the specified safety function.\par}
\end{table}

\begin{table}[p]
\ContinuedFloat
\centering
\caption[]{\textbf{Reliability-assessment modelling from a coin toss to an LLM service (continued).} \textbf{b}, Operational use, inference and claim validity.}
\label{tab:comparison-b}
\RelTableStyle
\begin{tabularx}{\linewidth}{@{}Z{0.85}Z{0.65}Z{1.05}Z{1.00}Z{1.10}>{\columncolor{RelTableLLM}}Z{1.35}@{}}
\toprule
\RelTableHeader
\midrule
\textbf{i.i.d. / dependence} & Plausible if the mechanism is stable and tosses reset & Independent draws from a stable OP, fixed configuration and specified resets; otherwise model dependence & Duplicates and user/acquisition clusters may induce dependence; class-mix shift affects representativeness & Adjacent miles, routes, weather, scenarios and versions can be correlated & Turns, users, task families, retrieval snapshots, context, routing and memory may induce clusters and state dependence \\[4pt]
\midrule
\textbf{Operational profile} & Known trial mechanism & Demand distribution justified for the plant and mission & Deployment input distribution; a public dataset is only a proxy, with sparse coverage of high-dimensional regions & ODD plus probabilities over scenarios, routes, weather, traffic and driver behaviour & Dynamic distribution over users, intents, languages, histories, contexts, retrieval and serving conditions \\[4pt]
\midrule
\textbf{Local probabilistic robustness} & Normally not applicable & Input tolerances and neighbouring states addressed through requirements and robustness analysis & Label-preserving perturbations around an input & Nearby scenario parameters, trajectories, sensors and behaviours & Conditional model $K_u$ for task-preserving wording, format, language, context order and valid retrieval variation within one modelled demand \\[4pt]
\midrule
\textbf{Estimator / uncertainty / evidence} & MLE or Bayesian estimate; exact or asymptotic interval & Rare-event bounds, Bayesian or conservative inference, stopping rules and non-test assurance evidence & Weighted or stratified estimate; sampling, label, shift and calibration uncertainty & Rate, survival or Bayesian model; exposure, reporting, simulation and transfer uncertainty & OP-weighted and hierarchical inference; sampling, oracle, output, dependence, $K_u$, rare-event and drift uncertainty \\[4pt]
\midrule
\textbf{Version / claim lifetime} & Stability is assumed & Bound to software, hardware, plant state and demand profile & Model, preprocessing, labels and user distribution can change & Full stack, maps, ODD, traffic, weather and reporting evolve & Model, prompt, decoding, retrieval, filters and routing can drift; judge and OP changes also affect the assessment \\[4pt]
\bottomrule
\end{tabularx}
\par\smallskip
{\fontsize{8}{10}\selectfont\RaggedRight
OP, operational profile; PFD, probability of failure on demand; ODD, operational design domain; DL, deep learning; NHPP, nonhomogeneous Poisson process; i.i.d., independent and identically distributed; MLE, maximum-likelihood estimate.\par}
\end{table}

\subsection{Claim and metric: specify exposure and acceptance requirements}

The first two rows concern the metric and statistical unit. \emph{On-demand systems} act on discrete demands, whereas \emph{continuously operating systems} remain active in controlling a process\citep{strigini_guidelines_1997}. The former are naturally assessed by failure probability per demand; the latter by failure occurrence rates or failure-free probabilities over stated intervals. For example, a classifier may be assessed per image, while autonomous vehicles may be assessed per scenario or trip, or by events per mile or hour\cite{zhao2025statistical}. These metrics are not interchangeable, and their statistical assumptions require justification \cite{strigini_guidelines_1997,zhao2020operational}.

For LLMs, a prompt--response pair may constitute one demand for a narrow classification-like service, whereas ``resolve one billing dispute'' may span several turns. The assessment must define task boundaries, reset conditions, and how escalation, abandonment or a turn limit affects the outcome. Other claims may concern specified failures per response or incidents per operating hour. The exposure unit and outcome definition must match the service being assessed.

A claim must also specify the required reliability level and the strength of evidence needed to support it\citep{strigini_guidelines_1997}. Failure probability describes service behaviour; confidence in its estimated bound concerns uncertainty in the assessment. Where an LLM is a component, its reliability requirement must be justified against system-level requirements, accounting for safeguards, recovery and component dependencies. Response accuracy alone cannot establish the reliability of a workflow involving retrieval, tools or human review.

\subsection{Failure and oracle: separate the outcome from its judgement}

The next two rows distinguish what constitutes failure from how it is observed. A coin's outcome is directly visible. Safety software is judged against a specified function, whereas classifiers depend on reference labels (e.g., provided by humans, which can be noisy) and autonomous vehicles may require event reconstruction and expert review. Thus, even before LLMs, an observed label need not be an error-free account of failure.

For LLMs, both decisions become more demanding. Factual correctness may have an objective answer, while acceptable tone in an email may depend partly on user requirements. Outcomes can also differ in severity: an awkward phrase, an unresolved request and materially incorrect advice should not automatically receive the same treatment. A binary failure definition remains possible, but must state which outcomes count as failures; graded loss and severity should remain visible where relevant.

The oracle is a separate measurement process. Human raters can disagree, user feedback is selective, and model judges can exhibit position, verbosity and self-preference biases.\citep{belz2020human,zheng2023judge} Changes in the rubric or rater population must not be mistaken for changes in service reliability. Observer-error, ordinal and hierarchical measurement models provide possible starting points for propagating these uncertainties into the claim.\citep{jacobs2021measurement,dawid1979observer}

Zhou et al. \citep{zhou2024reliable} distinguished correct, avoidant and incorrect responses across GPT, LLaMA and BLOOM model families. In their settings, instruction-adapted models were more often correct and more stable across prompts, but also avoided fewer questions and produced more plausible wrong answers. Human participants overlooked some errors, particularly on difficult questions. This illustrates two table entries: whether avoidance meets the service requirement, and whether the oracle detects failure accurately. An aggregate correctness score settles neither.

\subsection{Stochastic process and dependence: model how failures occur}

Rows five and six concern the outcome process and dependence between observations. Independent coin tosses and representative, reset software demands can support a Bernoulli model \cite{strigini_guidelines_1997}. Classifier evaluations can nevertheless contain duplicates or acquisition clusters, while adjacent vehicle miles share routes, weather and traffic. Independence is therefore an assumption to justify, not a property conferred by counting more tests.

LLMs add stochastic generation and interaction state. Earlier responses affect later prompts; requests can share users, incidents, retrieval snapshots, caches or serving routes. Binary outcomes may require a state-dependent model, while graded outcomes or incident counts require a different observation model. Positive dependence can make 2,000 responses contain substantially less information than 2,000 independent trials, producing overly narrow intervals if ignored \citep{salako2023unnecessity,salako2024correlated}.

Data collection should preserve conversation boundaries, order, reset policy and relevant cluster and service identifiers, subject to privacy constraints. Cluster-aware inference, hierarchical models, block bootstrap, hidden-state models and recurrent-event processes address different forms of dependence; none is universally appropriate \citep{min2022recurrent,chakherlou2026memory}. The process model must follow the chosen exposure and the mechanisms linking outcomes.

\subsection{Operational profile: represent the intended use}

The OP row asks \textit{which demands occur and how frequently}. The coin example has no user task mix. Safety software requires demand frequencies justified for its plant and mission \cite{strigini_guidelines_1997,musa1993op}; classifiers require a deployment input distribution; autonomous vehicles require probabilities within an operational design domain, not merely a list of possible scenarios \cite{zhao2025statistical}.

An LLM OP similarly needs frequencies, but over a potentially large combination of intents, languages, expertise, histories, document types, retrieval conditions and other relevant circumstances. A public dataset is only a proxy unless its relationship to that use is established. Deployment records are also incomplete: important task groups may be sparse, privacy limits personalization, and recorded feedback may be selective. In HIP-LLM's evaluation, GPT-4o scored 90.9\% and 55.2\% on two reading-comprehension test sets. Changing only the assumed mix of these tasks (i.e. OPs) substantially changed its inferred reliability \citep{chakherlou2026hip}. These OPs were illustrative, not measured.

User requirements and the raters assessing them must also be distinguished. PRISM documents variation in preferences and judgements across people and cultures \citep{kirk2024prism,casper2023rlhf}. The population whose service is being assessed need not coincide with the population recruited to judge its outputs. Both should be specified rather than silently represented by an average benchmark participant.

\subsection{Local probabilistic robustness: represent variation within a demand}

The local-robustness row concerns the resolution at which an OP represents use. At the chosen resolution, \emph{a modelled demand is the smallest unit of use distinguished by the model}. Its granularity depends on the service and reliability claim: a demand may involve classifying an image, answering a specified question or completing a bounded conversation. The OP records how often each modelled demand occurs, without necessarily distinguishing every concrete input that can realise it.

\begin{figure}
    \centering
    \includegraphics[width=1\linewidth]{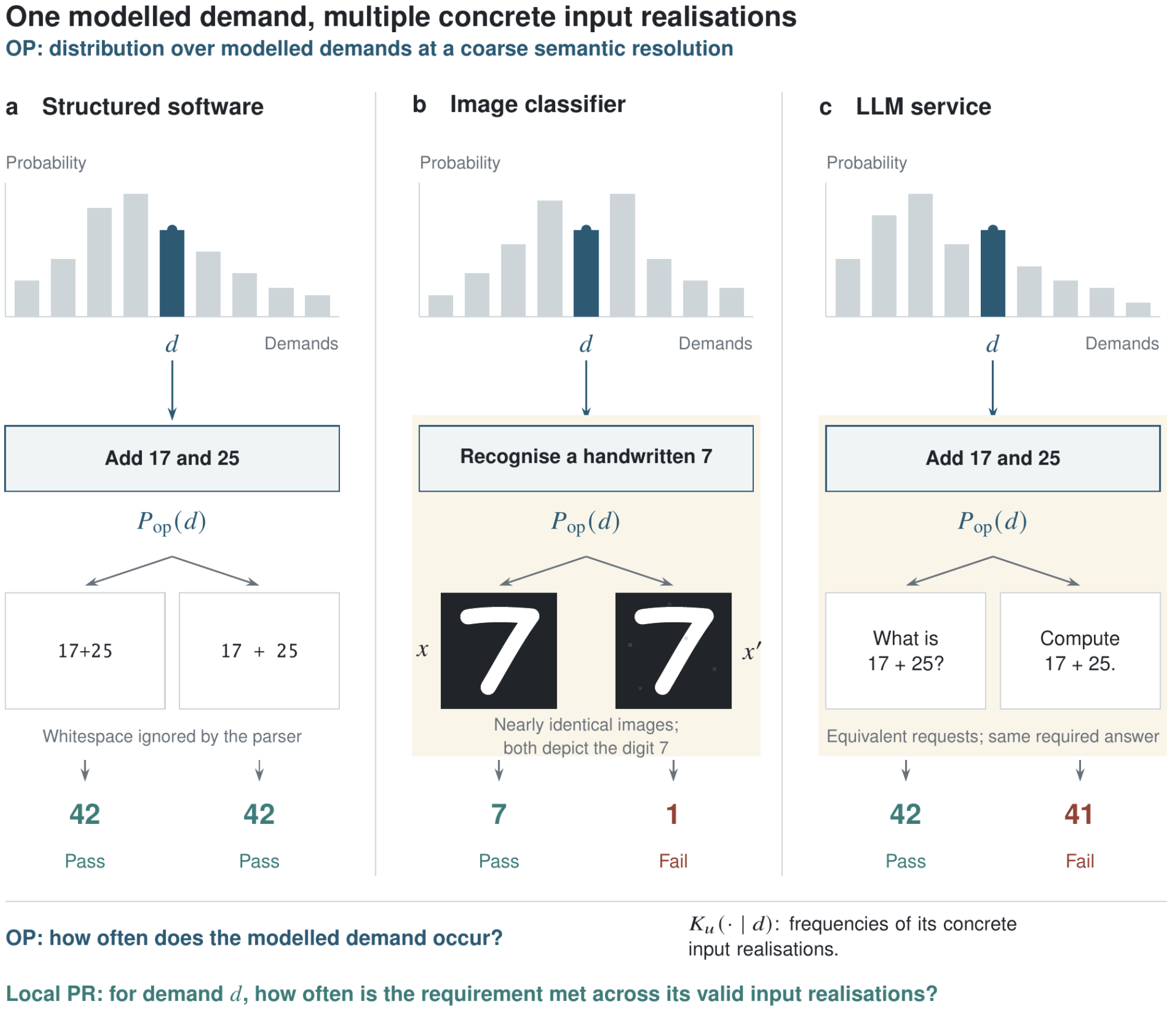}
    \caption{An illustrative example of why local probabilistic robustness matters in reliability assessment of AI models. The upper panels show schematic operational profiles (OPs) over modelled demands at a chosen semantic resolution. Each highlighted bar selects one demand \(d\), whose concrete input realisations are shown below. A calculator that ignores whitespace returns the same answer for equivalent inputs (a), whereas nearly identical images (b) or equivalent requests (c) may elicit different AI outputs. The OP specifies how often \(d\) occurs; \(K_u(\cdot\mid d)\) describes the frequencies of its valid input realisations. Local probabilistic robustness assesses the probability of meeting the requirement across these realisations.}
    \label{fig_OP_PR}
\end{figure}

Figure~\ref{fig_OP_PR} illustrates why this distinction matters. A calculator whose parser ignores whitespace treats ``17+25'' and ``17 + 25'' as the same calculation and returns the same result. Such representation equivalence can be explicitly enforced in traditional software (although it is not an inherent guarantee). For AI models, two images differing by a small, task-preserving pixel perturbation may appear indistinguishable to a person yet receive different classifications. Likewise, equivalent wordings of an LLM request may produce acceptable and incorrect answers. These are different input realisations of the same modelled demand, not necessarily different demands in the OP. Semantic equivalence therefore does not establish equivalent failure risk \citep{zhao2021classifiers,sclar2024sensitivity}.

Sclar and colleagues \citep{sclar2024sensitivity} reported differences of up to 76 percentage points in accuracy for LLaMA-2-13B across meaning-preserving prompt formats in few-shot settings, and showed that format choice could affect model comparisons. These results demonstrate variation within a modelled demand, but not how frequently it causes failure in operation. That requires evidence about which formats users actually encounter and how often.

Local probabilistic robustness \cite{webb2019statistical} makes this input variation explicit. Let $K_u(\cdot\mid d)$ describe how user population $u$ realises a modelled demand $d$ through plausible, task-preserving inputs. Local probabilistic robustness assesses the probability of meeting the requirement over these realisations, rather than merely reproducing the output for one canonical input. The OP supplies the demand frequencies; the conditional model supplies the frequencies of its input realisations. Reliability assessment combines these with the model's behaviour, including relevant generation randomness and state. Testing one canonical image or prompt cannot establish the risk across all its operationally relevant realisations \citep{zhao2021classifiers,webb2019statistical,zhao2026probabilistic}.

The challenge is to justify $K_u$: its variations must preserve the demand and acceptable-answer set, their frequencies must reflect intended use, and the resulting risk estimates must be checked against operational evidence. Small $L_p$ distance or embedding proximity alone does not establish demand preservation. Whenever an OP leaves behaviourally important input variation unresolved, that variation must be accounted for. If it is already captured by the OP and representative sampling, local probabilistic robustness makes an existing part of the assessment explicit; it must not be counted again as an independent risk factor.

\subsection{Estimator and uncertainty: state what the evidence supports}

The estimator row shows why a sample proportion is only a starting point. Under a justified independent-trial model, coin or binary demand outcomes support standard binomial inference. Safety-critical software may require rare-event bounds \cite{littlewood1993ultra,zhao2020operational}; classifier assessment may require weighting and label-error treatment; vehicle assessment may require exposure-based models and arguments connecting simulation to operation \cite{zhao2020operational,zhao2025statistical}.

LLM inference must match the preceding choices: the target OP, outcome measure, oracle, dependence structure and within-task variation. The resulting estimate or bound should state its confidence or credible level and which uncertainties it includes. More observations reduce sampling uncertainty under the model; they do not by themselves correct an unrepresentative OP, biased judge or inappropriate independence assumption.


Acceptance concerns whether the evidence supports a stated operational criterion, not merely whether the point estimate looks favourable. Design evidence, containment and monitoring can complement testing, provided their contributions and limitations remain explicit \citep{bishop2011formalism,nist2023airmf,iso25059}.

\subsection{Version and claim lifetime: define when reassessment is needed}

The final row asks when the assessment remains applicable. Even coin inference assumes a stable mechanism. Software claims depend on the assessed configuration and operating conditions \cite{littlewood2020replacement}; classifiers add preprocessing and distribution changes; vehicles add an evolving full stack and operating environment. Reliability evidence has conditions of validity in every column.

For an LLM service, the assessed configuration includes model weights, system prompts, decoding, retrieval, filters, caching and routing. Changes to this configuration must be distinguished from changes in the OP or the oracle: each can invalidate a different part of the assessment. Chen et al. \citep{chen2023behavior} documented substantial behavioural differences between ChatGPT service versions assessed only months apart. Work on distribution shift likewise explains why evidence from one user population or period need not transfer unchanged to another \citep{gama2014drift,ovadia2019shift}.

An operational claim should therefore identify the service and measurement versions, the use conditions it covers, and the changes that trigger reassessment. Lightweight monitoring between fuller acceptance campaigns offers a practical direction. The roadmap that follows develops these linked modelling needs into research questions and assessable outputs.

\section{A user-specific reliability roadmap}

Consider a retrieval-augmented customer-support LLM that reads a restricted account summary and policy documents, answers billing questions, recommends escalation, and cannot execute account actions. A meaningful candidate claim is:

\begin{tcolorbox}[
  colback=NatureWarm,colframe=NatureGold,boxrule=0pt,
  borderline west={2.4pt}{0pt}{NatureGold},arc=0pt,
  left=11pt,right=11pt,top=8pt,bottom=8pt]
\itshape 
\textbf{Reliability Claim}: For customers represented by the OP $P_u$, accept service version $v$ for one billing-dispute conversation only if the one-sided 95\% upper confidence bound on severity-3-or-higher failure probability is below 2.5\%. The decision applies within a stated validity window and change limits.
\end{tcolorbox}

\begin{figure}[H]
\centering
\includegraphics[width=1\linewidth]{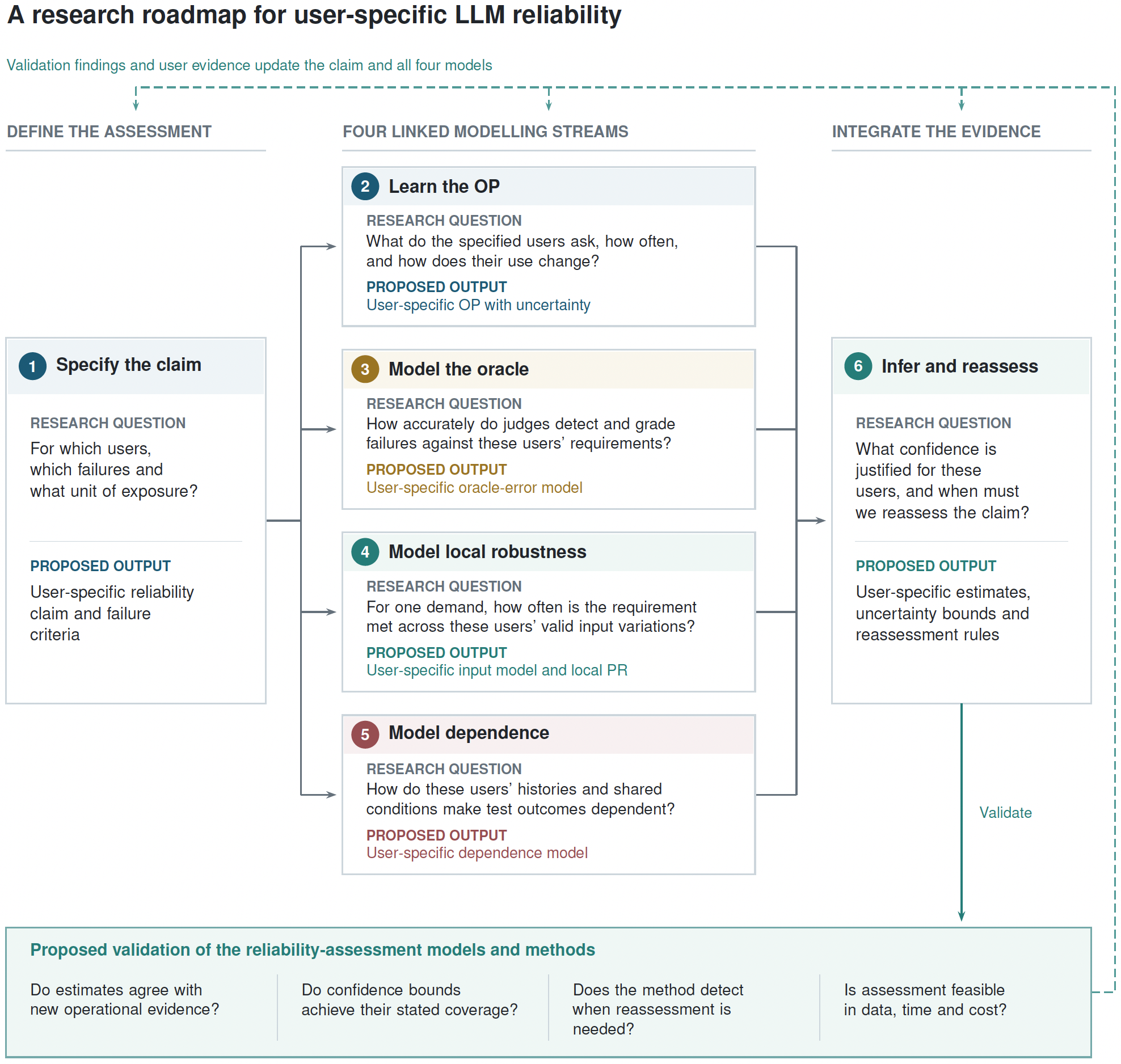}
\caption{\textbf{Six linked workstreams for user-specific LLM reliability.} Each workstream states a research question and a proposed output. A user-specific claim guides four linked modelling streams: the operational profile, oracle error, local probabilistic robustness, and dependence. Their integration supports reliability estimates, uncertainty bounds and reassessment rules for the specified users. The lower band proposes validation of the reliability-assessment models and methods, rather than another capability test of the LLM. Validation findings and operational evidence inform revisions throughout the roadmap.}
\label{fig_roadmap}
\end{figure}

This illustrative criterion makes the population, unit, severity and statistical decision explicit; its numerical threshold is hypothetical, not a deployment recommendation. The roadmap in Figure \ref{fig_roadmap} organizes the associated research into six linked workstreams. Operational evidence updates the OP and oracle and informs the next assessment. Each workstream needs validation against held-out operation and sensitivity to competing modelling choices.

\subsection{1. Define the user-specific claim and assessment unit}

The first research problem is to specify whose reliability is being assessed, what service they require, and what counts as one demand and one failure. In the running example, a conversation starts with a new billing-dispute request and ends with an adequate answer, escalation, abandonment or a predefined interaction limit. The specification must state whether escalation is acceptable, how unfinished conversations are classified, and what context or state is reset between conversations. Under the illustrative rubric, severity-3-or-higher failures might include materially incorrect billing guidance, disclosure of private account information or falsely reporting resolution. These are service-failure criteria, not a complete harm or threat model. The assessed service configuration includes the model, system prompt, decoding, retrieval, filters, permissions and routing. The rubric and judge are versioned separately: changing the measurement procedure can change measured reliability without changing the service.

The output is a claim specification recording the user population, requirements, assessment unit, failure criteria, acceptance threshold and assessed configuration. A central research question is whether these objects can be specified precisely enough for independent assessors to construct equivalent assessment units and interpret outcomes consistently. Without this agreement, different reliability estimates may simply answer different questions.

\subsection{2. Learn and update a user-specific OP}

The OP should distinguish modelled demands at a resolution that is relevant to the claim and supportable by available evidence. For the support service, demands might be distinguished by billing intent, language, customer circumstances and retrieval conditions. Operational records can estimate common demand frequencies; targeted collection can improve coverage of rare but consequential demands, provided the sampling design is accounted for. Hierarchical partial pooling can stabilize estimates for small user groups while retaining differences and uncertainty. Where precise frequencies are not defensible, sets or intervals of plausible OPs can represent the remaining uncertainty \citep{pietrantuono2020op,chakherlou2026hip}.

Key questions are how to choose a useful demand representation, estimate its frequencies under privacy constraints, and detect when users' patterns of use have changed. The output is a timestamped OP with uncertainty estimates, coverage gaps and update triggers. Its adequacy should be tested by whether it helps predict outcomes in subsequent operation. A practical assessment cycle can combine lightweight monitoring, periodic updates and fuller acceptance campaigns after material changes, rather than requiring an equally expensive campaign for every release.

\subsection{3. Model user-relevant failure and calibrate the oracle}

The third workstream connects the specified users' requirements to graded outcomes and an explicit measurement procedure. For the running example, expert reviewers can assess severity and task completion on a stratified subset, while rules or an LLM judge provide preliminary labels at scale. Latent-class, ordinal or hierarchical models offer possible ways to estimate judgement errors and differences between raters. These models require identifying assumptions and audited reference cases: agreement between judges alone cannot reveal a shared bias. Selectively recorded user feedback also requires a model of which outcomes are observed.

The research questions concern calibration and transfer. How much audited evidence is needed in each important demand group? When can judge-error estimates be shared across languages or user groups? How should changes to the rubric or judge be distinguished from changes in service reliability? The output is a versioned failure rubric and calibrated oracle model, supported by severity anchors, audited cases and uncertainty estimates. Reliability inference must propagate this measurement uncertainty rather than treat every recorded label as an error-free observation.

\subsection{4. Assess local probabilistic robustness}

For each modelled demand $d$, the conditional model $K_u(\cdot \mid d)$ should describe how the specified users realise that demand through concrete inputs. Relevant variation may include wording, formatting, language variety, attachment layout and alternative valid retrieved passages. The research problem is not simply to generate paraphrases, but to establish which variations preserve the demand and how frequently they occur. Historical user edits, naturally occurring alternative expressions and controlled elicitation can inform this model. Local probabilistic robustness then concerns how often the service meets the requirement across those realisations.

The output is a justified within-demand variation model and an assessment of local probabilistic robustness. Validation must examine both demand preservation and operational plausibility; embedding proximity alone establishes neither. Where frequencies remain uncertain, a conservative direction is to lower-bound local probabilistic robustness over a justified set of candidate variation distributions, rather than assume one known distribution \citep{wang2026nppr}. Competing representations should be compared through their ability to predict failure under held-out operational variation. Variation already captured by the OP and its sampling design must not be counted again.

\subsection{5. Model dependence and interaction state}

Local variation describes how a demand is realised; dependence concerns how outcomes are linked within and across executions. For the specified users, earlier turns can change later prompts, repeated conversations can share customer circumstances, and requests can share incidents, retrieval snapshots or serving conditions. These observations should not be assumed independent without justification. The research question is which dependencies materially affect the reliability estimate and the amount of information supplied by a test campaign.

Candidate approaches include cluster-aware inference for shared users or incidents, and hidden-state models for memory-dependent behaviour \citep{salako2024correlated,chakherlou2026memory}. The output is a justified process model and data-collection protocol recording relevant conversation boundaries, ordering, resets and service state. Its adequacy should be assessed through predictive checks and sensitivity to alternative dependence assumptions. Integration with workstream 4 must preserve any relationship between input variation, interaction history and service state; separating the workstreams does not assume that these objects are statistically independent.

\subsection{6. Infer user-specific reliability and define reassessment}

The final workstream combines the preceding models into a statistical assessment of the claim. Acceptance cases may be sampled directly from the OP or through a documented stratified or weighted design. Debug cases can also contribute, but reweighting requires coverage of the claimed operational demands and justified sampling probabilities. Weighting cannot replace observations from excluded regions. Diagnostics should examine coverage gaps, extreme weights and effective sample size alongside the dependence model. Candidate methods include weighted estimators, hierarchical inference, cluster-aware intervals, conservative Bayesian methods and sequential procedures. Their suitability follows from the claim and models developed in workstreams 1--5.

Suppose, purely for illustration, that the analysis estimates a severity-3-or-higher failure probability of 1.8\%, with a one-sided 95\% upper confidence bound of 2.9\%. Although the point estimate is below the hypothetical 2.5\% criterion, the upper bound is not; the evidence therefore does not support acceptance. These numbers illustrate a decision rule, not an experimental result. Research must establish how uncertainty in the OP, oracle, local variation and dependence propagates into such bounds, including under adaptive sampling and repeated assessment.

\begin{table}[t]
\centering
\caption{\textbf{Illustrative reliability assessment record.} The customer-support example connects the roadmap to one acceptance claim.}
\label{tab_assessment_example}
\small
\setlength{\tabcolsep}{8pt}
\renewcommand{\arraystretch}{1.12}
\begin{tabularx}{\linewidth}{@{}>{\raggedright\arraybackslash}p{0.23\linewidth}>{\raggedright\arraybackslash}X@{}}
\toprule
\textbf{Assessment element} & \textbf{Illustrative specification} \\
\midrule
\textbf{Users and service} & Customer population $u$ using billing-support service $v$, which retrieves approved information and recommends escalation but cannot change accounts. \\
\addlinespace[5pt]
\textbf{One demand} & One billing-dispute conversation, ending with an answer, escalation, abandonment or a predefined interaction limit. The rubric specifies how each endpoint is scored. \\
\addlinespace[5pt]
\textbf{Failure and severity} & Severity-3-or-higher failure: materially incorrect advice, private-account disclosure or falsely reporting resolution, assessed against explicit severity anchors. \\
\addlinespace[5pt]
\textbf{OP and local variation} & $P_u(d)$ gives demand frequencies; $K_u(\cdot\mid d)$ gives frequencies of valid wording, formatting and context variations within a demand. Both need operational justification. \\
\addlinespace[5pt]
\textbf{Measurement} & Expert-audited cases calibrate the rules or model judge. Judge error and rater differences enter the reliability assessment. \\
\addlinespace[5pt]
\textbf{Dependence and reset} & Reset conversation context between tasks. Record shared customer, incident, retrieval and serving conditions, and assess any remaining outcome dependence. \\
\addlinespace[5pt]
\textbf{Metric and requirement} & The probability $p_{\geq 3}$ of a severity-3-or-higher failure per conversation under the specified use. Acceptance requires its one-sided 95\% upper confidence bound to be below 2.5\%. \\
\addlinespace[5pt]
\textbf{Illustrative decision} & $\widehat p_{\geq 3}=1.8\%$, but $U_{0.95}=2.9\%>2.5\%$: evidence does \emph{not} support acceptance. This does not prove that the true probability exceeds the threshold. \\
\addlinespace[5pt]
\textbf{Validity and reassessment} & Record service, OP and oracle versions, a review date and material-change triggers. Changes in users, retrieval, prompts, scoring or serving may require reassessment. \\
\bottomrule
\end{tabularx}
\par\vspace{5pt}
\begin{minipage}{\linewidth}
\footnotesize
\textit{Illustration only.} Requirements and numbers are hypothetical. Constructing and validating the assessment remains research tasks.
\end{minipage}
\end{table}

The output, as an illustrative example shown in Table~\ref{tab_assessment_example} is a versioned decision record connecting the claim, observations, models, estimator and uncertainty bound. It should specify a review date and the changes that require earlier reassessment, including changes to the service, user population, OP, rubric or judge. Validation should test predictions against subsequent operation and examine interval coverage in controlled studies with known risks. The goal is a traceable, renewable reliability claim for the specified users, not a permanent score attached to a model name.




\section{Limits of the position}

Four boundaries should remain explicit. First, the roadmap addresses reliability under a stated operating population. Safety additionally requires harm, exposure and mitigation analysis; security requires threat and asset models \citep{avizienis2004taxonomy,bloomfieldRushbyWhereAIWrong2024}. Counting harmful outputs as failures can contribute evidence to both, but a low average failure probability does not by itself establish either property. Adaptive attackers can choose inputs in response to defences, so their risk need not follow the operational sampling model. Adversarial evaluation and appropriate worst-case or game-theoretic reasoning should support an accompanying security claim.

Second, this article focuses on LLM services rather than full AI-agent systems. Although the roadmap’s principles are generally applicable to agents, extending them requires additional system-level modelling. The assessment unit may be a complete mission trajectory, with explicit termination and reset conditions. The OP must account for environmental states, tool availability, permissions and human intervention. Failure assessment must consider intermediate actions and unacceptable side effects, even when the final response appears correct. Earlier actions can change subsequent conditions and failure risk, while tools, memory and external services can change independently of the base model. Evidence from simulation, replay and deployment also requires justification of its relevance to intended use. These extensions introduce further challenges of cost, reproducibility and operational relevance identified in agent evaluation \citep{kapoor2025agents,wang2024agents}. A full agent-level reliability framework therefore remains a further research task.

Third, some components may not be identifiable from available data. Sparse user groups, privacy restrictions, selective feedback and correlated judge error can leave multiple reliability models consistent with the observations. The appropriate output may then be a conservative bound or a statement that the claim is unsupported, not a forced point estimate.

Last, the article proposes a research agenda; the linked framework has not been empirically validated as a whole. The published studies motivate individual problems within their tested settings, rather than validating the proposed framework. The customer-support example and its numerical criterion illustrate modelling choices and decision logic, not new empirical results. Future work must establish whether the proposed OP and local-variation models predict deployment outcomes, whether oracle calibration transfers, and whether the resulting intervals have adequate coverage under realistic dependence and drift. Service changes require reassessment, and continuous monitoring needs rules that account for repeated decisions. The contribution is a coherent set of reliability questions and validation targets.

\section{Outlook}

The central question is not ``how good is this LLM in general?'' It is how often a specified user population will experience a specified failure, over a stated demand or exposure, for a frozen service version, with stated uncertainty. Traditional software reliability provides the disciplined chain from testing purpose to OP, failure process, estimator, evidence and bounded claim. It also explains why bug counts and benchmark averages cannot answer that question by themselves.

LLMs make every link harder: tasks are semantic and sequential, OPs are user-specific and dynamic, failure is graded, oracles are noisy, outcomes are dependent and operationally equivalent inputs can induce different behaviour. Treating probabilistic local robustness as the within-task layer of OP modelling identifies how to represent the variation hidden by coarse task categories. The research programme is concrete: define claim units, learn dynamic OPs, calibrate noisy oracles, validate local-variation models, represent dependence, design acceptance sampling and control the evidence lifecycle.

Success should be measured by whether assessors can reconstruct the same bounded claim, trace its evidence, quantify what the data do not establish and recognize when reassessment is needed. Benchmarks remain valuable within this evaluation landscape. Connecting them and deployment observations to delivered, user-specific reliability would make their implications for operational decisions more precise.

\section*{Competing interests}
The author declares no competing interests.

\section*{Use of AI-assisted tools}
Generative AI tools assisted language editing, document organization and bibliography preparation for this paper. The author remains responsible for verifying every claim and reference. 

\begingroup
\small
\setlength{\bibsep}{3.0pt}
\bibliographystyle{naturemag}
\bibliography{ref_nature}
\endgroup

\end{document}